\documentclass[aps,prl,twocolumn,superscriptaddress,preprintnumbers]{revtex4}

\usepackage{epsf,epsfig,amsmath,amssymb,amsfonts}
\usepackage{color}
\usepackage[thinlines]{easytable}
\usepackage{slashed, comment}
\usepackage{hyperref}

\PassOptionsToPackage{table}{xcolor}
\hypersetup{ linktoc=all,
    colorlinks, linkcolor={blue}, 
    citecolor={red}, urlcolor={darkpink}
}

\graphicspath{{Images/}}

\definecolor{light gray}{RGB}{220,220,220}
\definecolor{dark purple}{RGB}{108,0,217}
\definecolor{pink}{RGB}{190,20,100}
\definecolor{orang}{RGB}{193,63,0}
\definecolor{green}{RGB}{11,98,17}
\definecolor{darkpink}{RGB}{153,0,76}
\definecolor{bluegreen}{RGB}{0,102,102}
\definecolor{greenlagan}{RGB}{0,102,0}
\definecolor{redgreen}{RGB}{102,102,0}
\definecolor{Redgreen}{RGB}{153,76,0}
\definecolor{vividviolet}{rgb}{0.62, 0.0, 1.0}
\definecolor{amaranth}{rgb}{0.9, 0.17, 0.31}
\definecolor{palatinateblue}{rgb}{0.15, 0.23, 0.89}
\definecolor{brightpink}{rgb}{1.0, 0.0, 0.5}
\definecolor{cornflowerblue}{rgb}{0.39, 0.58, 0.93}
\definecolor{deepcarminepink}{rgb}{0.94, 0.19, 0.22}
\definecolor{radicalred}{rgb}{1.0, 0.21, 0.37}

\newsavebox{\ns}
\newsavebox{\dbrane}

\def\be{\begin{equation}}
\def\ee{\end{equation}}
\def\bea{\begin{eqnarray}}
\def\eea{\end{eqnarray}}

\def\Dslash{\,\,{\raise.15ex\hbox{/}\mkern-12mu D}}
\def\Dbarslash{\,\,{\raise.15ex\hbox{/}\mkern-12mu {\bar D}}}
\def\delslash{\,\,{\raise.15ex\hbox{/}\mkern-9mu \partial}}
\def\delbarslash{\,\,{\raise.15ex\hbox{/}\mkern-9mu {\bar\partial}}}
\def\pslash{\,\,{\raise.15ex\hbox{/}\mkern-9mu p}}
\def\calDslash{\,\,{\raise.15ex\hbox{/}\mkern-12mu {\cal D}}}

\newcommand\diff{\mbox{d}}

\newcommand{\nn}{\nonumber \\}

\newcommand{\dd}{\diff}

\allowdisplaybreaks

\begin{document}

\title{Hubble Sinks In The Low-Redshift Swampland}

\vskip 1 cm

\author{A. Banerjee}\email{aritra.banerjee@apctp.org}
\affiliation{Asia Pacific Center for Theoretical Physics, Postech, Pohang 37673, Korea}
\author{H. Cai}\email{haiying.cai@apctp.org }
\affiliation{Asia Pacific Center for Theoretical Physics, Postech, Pohang 37673, Korea}
\author{L. Heisenberg}\email{laviniah@phys.ethz.ch}
\affiliation{Institute for Theoretical Physics, ETH Zurich, Wolfgang-Pauli-Strasse 27, 8093, Zurich, Switzerland}
 \author{E. \'O Colg\'ain}\email{ocolgain.eoin@apctp.org}
 \affiliation{Asia Pacific Center for Theoretical Physics, Postech, Pohang 37673, Korea}
\affiliation{Department of Physics, Postech, Pohang 37673, Korea}
\author{M. M. Sheikh-Jabbari}\email{jabbari@theory.ipm.ac.ir}
\affiliation{School of Physics, Institute for Research in Fundamental Sciences (IPM), P.O.Box 19395-5531, Tehran, Iran}
\affiliation{The Abdus Salam ICTP, Strada Costiera 11, 34151, Trieste, Italy}
\author{T. Yang}\email{tao.yang@apctp.org}
\affiliation{Asia Pacific Center for Theoretical Physics, Postech, Pohang 37673, Korea}

\begin{abstract}
\noindent 
Local determinations of the Hubble constant $H_0$ favour a higher value than Planck based on CMB and $\Lambda$CDM. Through a model-independent expansion, we show that low redshift ($z \lesssim 0.7$) data comprising baryon acoustic oscillations (BAO), cosmic chronometers and Type Ia supernovae has a preference for Quintessence models that lower $H_0$ relative to $\Lambda$CDM. In addition, we confirm that an exponential coupling to dark matter cannot alter this conclusion in the same redshift range. Our results leave open the possibility that a coupling in the matter-dominated epoch, potentially even in the dark ages, may yet save $H_0$ from sinking in the string theory Swampland. 

\end{abstract}

\maketitle

\setcounter{equation}{0}

\section{Introduction} \label{Introduction}
Last century Allan Sandage framed cosmology as the search for two numbers \cite{Sandage}. 50 years later, it is widely accepted that the deceleration parameter $q_0$ is negative \cite{Riess:1998cb, Perlmutter:1998np}, and the Hubble constant $H_0$ is known to within $10 \, \%$ \cite{Aghanim:2018eyx, Riess:2019cxk, Wong:2019kwg, Freedman:2019jwv, Pesce:2020xfe}, making it one of the best measured quantities in late-time cosmology. At the lower end of this $H_0$ window, one finds the Planck determination based on the cosmological model $\Lambda$CDM and the Cosmic Microwave Background (CMB) \cite{Aghanim:2018eyx}. In contradistinction, a host of local determinations favour higher values \cite{Riess:2019cxk, Wong:2019kwg, Freedman:2019jwv, Pesce:2020xfe}, resulting in Hubble tension  \cite{Verde:2019ivm}, an intriguing discrepancy between early and late Universe determinations of $H_0$. 

The convergence of different experiments suggests that a high local $H_0$ is legit (to lose one experiment, may be regarded as a misfortune; to lose all looks like carelessness). One logical possibility, advocated by the Swampland programme within string theory \cite{Agrawal:2018own} is that we replace $\Lambda$ with Quintessence \cite{Copeland:2006wr, Tsujikawa:2013fta}, the simplest alternative in effective field theory (EFT). Against this backdrop, our goal is to ascertain if $H_0$ is raised or lowered relative to $\Lambda$CDM by the Swampland.  

To date the Swampland \cite{Vafa:2005ui} has led to an intriguing web of conjectures (see \cite{Brennan:2017rbf, Palti:2019pca} for reviews), which remarkably impinge on the real world. The most consequential precludes de Sitter vacua \cite{Obied:2018sgi} and is in conflict with $\Lambda$CDM \cite{Agrawal:2018own}. Subsequent to-and-fro discussion led to a refinement  \cite{Garg:2018reu, Ooguri:2018wrx}  (also \cite{Andriot:2018wzk}), thus allowing the conjecture to co-exist with the Higgs' potential \cite{Denef:2018etk}. Nevertheless, the de Sitter conjecture is constrained by cosmological data \cite{Heisenberg:2018yae, Akrami:2018ylq, Raveri:2018ddi} and it was highlighted early on that Hubble tension may be an issue \cite{Colgain:2018wgk}. 

Here, we return to the Hubble tension thread within the context of generic Quintessence models \cite{Copeland:2006wr, Tsujikawa:2013fta}. Recall that ``Quintessence" may be defined as a canonically coupled scalar with scalar potential $V> 0$ and an equation of state (EoS) for dark energy that is bounded below $w \geq -1$. However, the latter can be relaxed within EFT by coupling dark matter to dark energy through ``coupled Quintessence"  \cite{Wetterich:1994bg,Amendola:1999er, Das:2005yj}. While it is known that a coupling allows one to effectively reduce $w$ and hence to raise $H_0$ \cite{DiValentino:2016hlg, DiValentino:2019jae}, in contrast, specific uncoupled Quintessence models lower $H_0$ \cite{Vagnozzi:2019ezj, Alestas:2020mvb, Vagnozzi:2018jhn, Colgain:2019joh} (see also Table I \cite{Agrawal:2019dlm}) \footnote{The CPL model \cite{Chevallier:2000qy, Linder:2002et} clearly works well as a model-independent approach to dark energy when confronting observations. However, since the equation of state $w(z)$ is a quantity derived from a potential, the fact that the CPL expansion is truncated at linear order in $1-a$ means that the corresponding potential will be constrained. It is not a generic Quintessence model, especially when one fits data from CMB $z \approx 1100$. In particular, the Quintessence potential $V(\phi) = V_1 e^{-\lambda_1 \phi_1} + V_2 e^{-\lambda_2 \phi_2}$, $\lambda_1 \gg 1, \lambda_2 \lesssim 1$ does not admit a CPL prescription.}, thus implying that $H_0$ is in the Swampland. In this letter we show that this is more generally true. 

In contrast to models with fixed potentials, e. g. \cite{vandeBruck:2017idm, vandeBruck:2019vzd, Miranda:2017rdk, Gomez-Valent:2020mqn, Ruchika:2020avj}, here we exploit perturbation at low redshift to work in a model-independent way. In short, we construct a large class of ``bottom-up" potentials, which we confront with direct measurements of the Hubble parameter $H(z)$ inferred {from cosmic chronometers \cite{CC1, CC2, CC3, CC4, CC5, CC6, CC7} and BAO \cite{Alam:2016hwk} \footnote{{Concretely, our Taylor expansion has only enough parameters to approximate the angular diameter distance $D_{A}(z)$ data to $z \approx 0.3$, so we can only use the $D_{H}(z) = c/H(z)$ data and its covariance matrix. This is not a problem for the isotropic BAO at $z = 0.106$ and $z = 0.15$.}}}. In addition, we include {low-$z$} BAO measurements by 6dF \cite{Beutler:2011hx} {($z= 0.106$)} and SDSS-MGS surveys \cite{Ross:2014qpa} {($z=0.15$)}, and determinations of $E(z) \equiv H(z)/H_0$ from type Ia supernovae \cite{Riess:2017lxs} {(derived from \cite{Scolnic:2017caz})}. Finally, we allow for an exponential potential, which is the dark matter-Quintessence coupling of choice in string theory. 

Our findings are simply expressed. Quintessence models  prefer a lower $H_0$ value than $\Lambda$CDM. Moreover, neither a $H_0$ prior \cite{Riess:2019cxk} nor an exponential coupling to dark matter - provided it is constrained by galaxy warps \cite{Desmond:2018euk, Desmond:2018sdy} - can change the conclusion in the redshift range. Models where the coupling is turned on earlier are still viable, but only at higher redshifts, namely within the matter-dominated era ($ 0.4 \lesssim z$).  Furthermore, if the coupling is turned on in the dark ages beyond $z \approx 6$ \cite{Agrawal:2019dlm}, a proper test of such a scenario may rest upon future developments in 21-cm cosmology \cite{Pritchard:2011xb, Bowman:2018yin}. Nevertheless, this loophole aside, low-redshift observations place the de Sitter conjecture \cite{Obied:2018sgi} at odds with local determinations of $H_0$, since the class of models championed by the conjecture only exacerbates Hubble tension. 

\section{Set-up} 
We consider flat FLRW spacetime and the following coupled Quintessence equations: 
\bea
\label{friedmann} H^2 &=& \frac{1}{3} \left( V + \frac{1}{2} \dot{\phi}^2 + f \frac{\rho_c}{a^3} + \frac{\rho_b}{a^3}\right), \\
\label{scalar} 0 &=& \ddot{\phi} + 3 H \dot{\phi} + \partial_{\phi} V + \frac{\rho_c}{a^3} \partial_{\phi} f, 
\eea
where dots denote time derivatives, $a$ is the scale factor and $H \equiv \frac{\dot{a}}{a}$ is the Hubble parameter. In addition, $f(\phi)$ is a coupling between the dark matter density $\rho_c$ and the Quintessence field $\phi$, while $\rho_b$ denotes the baryonic matter density. Observe that setting $f = 1$, we recover uncoupled Quintessence with matter density $\rho_m = \rho_c + \rho_b$. These equations allow an effective EoS $w_{\textrm{eff}} < -1$ when $f < 1$ \cite{Das:2005yj}. Concretely, we consider $f = e^{-C (\phi- \phi_0)}$ where $\phi_0$ and $C \geq 0$ are constant. Finally, observe that we have set $M_{p} =1$ for simplicity. 

We emphasise that the basic ingredients of  equations (\ref{friedmann}) and (\ref{scalar}), i. e. a canonically coupled scalar,  a potential and an exponential coupling are ubiquitous in low-energy effective descriptions of string theory, e. g. \cite{Brax:1999gp, Anchordoqui:2019amx,Anchordoqui:2020sqo, Chiang:2018jdg,DiValentino:2019exe, Ferrara:2019tmu,Montefalcone:2020vlu},  thus, the basic building blocks of the Swampland are in place. Now, to solve these equations, we recall the usual definition of the scale factor in terms of redshift $a  = (1+z)^{-1}$, normalised so that $a =1$ today. Using the chain rule, one establishes that 
${\dd}/{\dd t} = - (1+z) H {\dd}/{\dd z}$,  
and it is easy to recast the Friedmann (\ref{friedmann}) and scalar equation (\ref{scalar}) in terms of redshift. We consider the following expansion for the scalar, 
\be
\phi - \phi_0 = \alpha \, z + \beta \, z^2 + \gamma \, z^3 + \dots, 
\ee
around its value today $\phi_0$ at $z=0$, where $\alpha, \beta $ and $\gamma$ are constants. At small $z$, $\phi - \phi_0$ is small, and we can further expand the potential: 
\be
V = V_0 + V_0' (\phi- \phi_0 ) + \frac{1}{2} V_0'' (\phi - \phi_0)^2 + \dots, 
\ee
where we have defined $V_0 \equiv V(\phi_0), V_0' \equiv V'(\phi_0)$, etc. The Hubble parameter to third order in redshift is 
\be
\label{hubble}
H = H_0 ( 1+ h_1 z + h_2 z^2 + h_3 z^3 + \dots), 
\ee
where $h_i$ may be expressed in terms of the parameters as,
\begin{widetext} 
\bea
\label{h}
 h_ 1 &=& \frac{1}{2} \alpha^2 + \frac{3}{2} \Omega_{m0}, \qquad h_2 = \frac{1}{8} \alpha^4 + \frac{1}{4} \alpha^2 +  \alpha \beta  - \frac{3}{4} \alpha C \Omega_{c 0} +  \frac{3}{8} \Omega_{m0} (4 - 3 \Omega_{m0}), \nn
 h_3 &=& \frac{1}{48} \alpha^6 + \frac{1}{16} \alpha^4 ( \Omega_{m 0} + 2) +  \alpha \gamma  + \frac{1}{16} \alpha^2 \Omega_{m 0} \left( 9 \Omega_{m 0} - 2 \right) + \frac{1}{2} \alpha \beta \left( \alpha^2 + \Omega_{m 0} +\frac{4}{3} \right) + \frac{2}{3} \beta^2 \nn 
&+& \frac{1}{16} \Omega_{m 0} \left( 8 - 36 \Omega_{m 0} + 27 \Omega_{m 0}^2 \right) + \left( \frac{1}{8} \alpha^3  + \frac{9}{8} \alpha \Omega_{m 0}  - \alpha - \frac{1}{2} \beta \right) { \Omega_{c 0}} C + \frac{\alpha^2}{4} C^2 \Omega_{c 0}. 
\eea
\end{widetext}
Evidently, when $\alpha = \beta = \gamma = 0$, we recover the $\Lambda$CDM cosmology, while at $z=0$, we have $H(z=0) = H_0$.  
Observe  that we have employed the usual definitions $\Omega_{c 0} = {\rho_c}/(3 H_0^2), \quad \Omega_{b 0} = {\rho_b}/{(3 H_0^2)}, \quad \Omega_{m 0} = \Omega_{b 0} + \Omega_{c 0}$, 
to define the baryonic $\Omega_{b 0}$ and dark matter density $\Omega_{c 0}$ today.

Note that $\alpha = \beta = \gamma = 0$ is our reference model. Since we are working perturbatively around $z=0$, we first establish the range of validity of the expansion. To see this, we note that when $\Omega_{m 0} \approx 0.3$, namely the canonical Planck value \cite{Aghanim:2018eyx}, the corresponding values for $h_i$ are 
$h_1 \approx 0.45, \quad h_2 \approx 0.35, \quad h_3 \approx -0.007$. Concretely, if we demand that our expansion is within $1 \, \%$ of the exact analytic result, $E(z) = \sqrt{1 -\Omega_{m 0} + \Omega_{m 0} (1+z)^3}$,  then the linear, quadratic and cubic approximations are good to  $z \approx 0.18, z \approx 0.76$ and $ z \approx 0.82$, respectively. 

The potential and its derivatives at $z=0$ can be expressed in terms of $\alpha, \beta, \gamma$: 
\bea
\label{V0} V_0 H_0^{-2} &=& 3 \Omega_{\phi 0} - \frac{1}{2} \alpha^2, \\
\label{V1} V_0' H_0^{-2} &=& - \alpha (h_1 -2) - 2 \beta + 3 \Omega_{c 0} C, \\
\label{V2} V_0'' H_0^{-2} &=& - (h_1^2 -2 h_1 + 2 h_2 - 2) - 6 \frac{\beta h_1}{\alpha} \nn 
&-& \frac{6 \gamma}{\alpha}  - 3 \Omega_{c 0} \left( C^2 -3 \frac{C}{\alpha} \right), 
\eea
where for a given coupling $C$, we have a one-to-one correspondence between $(\alpha, \beta, \gamma)$ and $(V_0, V_0', V_0'')$. Note also that $\Omega_{\phi 0}$ is the dark energy density today  $\Omega_{\phi 0} = 1 - \Omega_{m 0}$. The effective EoS can easily be worked out, 
\begin{widetext} 
\bea
\label{w}
w_{\textrm{eff}} &=&  -1 + \frac{\alpha^2}{3 \Omega_{\phi 0}} + \frac{z}{\Omega_{\phi 0}^2} \biggl[ \frac{\alpha^4}{3} ( \Omega_{\phi 0} -1) + \frac{\alpha^2}{3} \Omega_{\phi 0} (5 - 3 \Omega_{\phi 0} ) + \frac{4}{3} \alpha \beta \Omega_{\phi 0} -  \alpha C \Omega_{c0} \Omega_{\phi 0} \biggr]\nn 
&+& \frac{z^2}{\Omega_{\phi 0}^3} \biggl[  \frac{\alpha^6}{6} ( \Omega_{\phi 0}^2 - 3 \Omega_{\phi 0} +2 ) + \frac{\alpha^4}{6} \Omega_{\phi 0} (17 \Omega_{\phi 0} -14 -3 \Omega_{\phi 0}^2) + 2 \alpha^3 \beta \Omega_{\phi 0} ( \Omega_{\phi 0} - 1 )   
+  \frac{\alpha^2}{3} \Omega_{\phi 0}^2 (10- 9 \Omega_{\phi 0})  \nn
&+& \frac{4}{3} \alpha \beta \Omega_{\phi 0}^2 (5 - 3 \Omega_{\phi 0}) + \frac{4}{3} \beta^2 \Omega_{\phi 0}^2 + 2 \alpha \gamma \Omega_{\phi 0}^2  
- C \Omega_{c 0} \Omega_{\phi 0} \left( \frac{1}{2} \alpha^3 ( \Omega_{\phi 0} -3)  + (3 \alpha + \beta)  \Omega_{\phi 0} \right) + \frac{1}{2} \alpha^2 C^2 \Omega_{c 0} \Omega_{\phi 0}^2 \biggr] \nn 
&+& \dots
\eea

\end{widetext}

Interestingly, regardless of the value of $C$, $w_{\textrm{eff}} > -1$ at $z=0$ and a crossing into the phantom regime ($w_{\textrm{eff}} < -1$) happens only at higher redshift.  Note that when $C=0$, we simply denote $w_{\textrm{eff}}$ as $w$. 

\section{Methodology} 
We begin by generating $10^6$ triples $(\alpha, \beta, \gamma)$, in a normal distribution about $(0,0,0)$ with a uniform standard deviation, which we take to be $\sigma = 0.1$ \footnote{Increasing this value or adopting a uniform distribution does not change our results.}. We impose a conservative redshift cut-off, $z_{\textrm{max}} = 0.7$, and impose
\be
\label{cut1}
| \alpha | \gtrsim  z_{\textrm{max}} |\beta|, \quad | \beta | \gtrsim  z_{\textrm{max}} |\gamma |, \quad | \phi - \phi_0| \lesssim 1, 
\ee
to ensure perturbation makes sense, i. e. higher order numbers are smaller. Note that these cuts can be implemented without establishing a best-fit value of $\Omega_{m0}$ and we are not imposing slow-roll. 

\begin{figure}[htb]
\begin{center} 
    \includegraphics[scale=0.6]{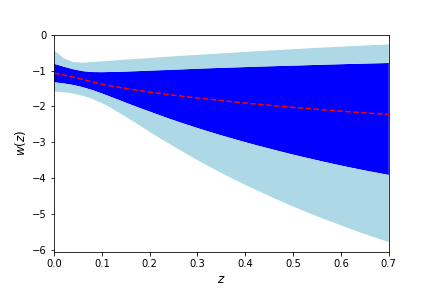}
   \caption{Bounds on the CPL parametrisation \cite{Chevallier:2000qy, Linder:2002et}, $w(z) = w_0 + w_a z/(1+z)$, which are constructed directly from the chains of an MCMC exploration of cosmic chronometer, BAO and supernovae data restricted to $z \leq 0.7$.}
 \end{center}
\end{figure}

For the triples surviving these primary cuts, we perform two-parameter fits of the data \cite{CC1, CC2, CC3, CC4, CC5, CC6, CC7, Alam:2016hwk, Beutler:2011hx, Ross:2014qpa,Riess:2017lxs} \footnote{For isotropic BAO, the expansion $D_{V}(z) = \frac{c}{H_0} (z - \frac{2}{3} h_1 z^2 + (\frac{19}{36} h_1^2-\frac{5}{9} h_2) z^3 + \dots )$ suffices.} and demand that 
\bea
\label{cut2}
V_0 & \gtrsim& |V_0'\cdot (\phi -\phi_0) | \quad |V'| \gtrsim \frac{1}{2} | V_0'' \cdot (\phi - \phi_0) |, 
\eea
once again to ensure a valid expansion. In addition, we impose the nominal restriction $\Omega_{m 0} \gtrsim 0.2$, which is in line with weak lensing results \cite{Abbott:2018xao, Joudaki:2019pmv}, and impose the $2 \, \sigma$ constraints from FIG. 1, which we have inferred from a Markov Chain Monte Carlo (MCMC) exploration of cosmic chronometer, BAO and Pantheon data below $z < 0.7$. Observe that in contrast to \cite{Heisenberg:2018yae}, our bounds come from low redshift data, which means they are less constraining, and for uncoupled models, we impose $w(z) > -1$. 

{It is worth emphasising that our expansion is expected to be valid to $z \approx 0.7$, but if we include distance moduli or angular diameter distance measurements, since they are integrated quantities, the approximation holds only for $z \lesssim 0.3$. This allows us to easily accommodate isotropic BAO at $z=0.106$ \cite{Beutler:2011hx} and $z = 0.15$ \cite{Ross:2014qpa}, but for higher redshift BAO we are restricted to { $D_{H} (z) \equiv c/H(z)$ and its covariance matrix}. For this reason, there is extra information in FIG 1. and this leads to additional constraints. Finally, since we have a large parameter space, e.g. $(H_0, \Omega_{m0}, \alpha, \beta, \gamma, C)$, one can expect any best-fit values to be within $1\, \sigma$ of $\Lambda$CDM, which ultimately favours the standard model. However, since our interest here is exploring the effect on $H_0$ of changing the underlying potential (theory), we simply scan over $\alpha, \beta, \gamma$ and $C$.}

Once the cuts are imposed, we are left with a large class of  ``bottom-up" Quintessence models, which we further divide into cosmologies where $H_0$ increases/decreases relative to the $\Lambda$CDM model, i. e. $\alpha = \beta = \gamma = 0$. For reference, we record the best-fit values  in this case, 
\bea
H_0 &=& {68.89^{+1.20}_{-1.15} } \textrm{ km s}^{-1} \textrm{ Mpc}^{-1}, \nn
\Omega_{m0} &=& 0.30^{+0.02}_{-0.02}. 
\eea
Note that the values are {pretty consistent} with Planck \cite{Aghanim:2018eyx} and due to our inflated errors, this represents {a $\sim 3.3 \, \sigma$ discrepancy} with the highest local $H_0$ determination based on cepheid-calibrated supernovae \cite{Riess:2019cxk}.  

Once the coupling is introduced, the dark matter particles are subject to a long-range attractive force with an effective Newton's constant $G_{\textrm{eff}} = G_N (1+ \frac{1}{2} C^2)$. While the CMB constrains the conformal coupling to be $C \lesssim 0.1$ \cite{vandeBruck:2017idm, Miranda:2017rdk}, if the coupling is turned on at late-times, e.g. \cite{Agrawal:2019dlm}, this constraint can be evaded. Nevertheless, stronger constraints exist for galaxies up to redshift $z \sim 0.05$ \cite{Desmond:2018euk, Desmond:2018sdy}, whereby $C \lesssim 0.05$. This number will be constrained further in future, but here we settle on the value $C = 0.1$. {This relaxes the constraint of \cite{Desmond:2018euk, Desmond:2018sdy} slightly, but still preserves its integrity as a strong observational bound.} Finally, since baryonic matter is uncoupled, we instead adopt the Planck prior $\Omega_{b0} = 0.05$ and only fit $H_0$ and $\Omega_{c0}$. 

\begin{figure}[htb]
\begin{center} 
    \includegraphics[scale=0.6]{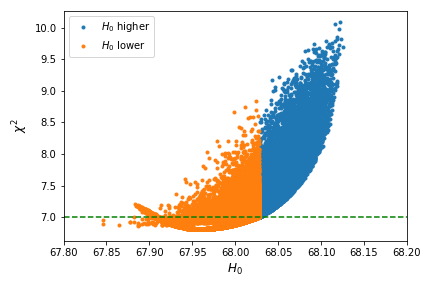}
   \caption{Distribution of our {87,675} models, which survive our cuts, as a function of $\chi^2$ versus $H_0$. Dashed green line highlights the $\chi^2$ for $\Lambda$CDM and all models that raise $H_0$ also increase the $\chi^2$ value. In contrast {53,744} models lower both $H_0$ and $\chi^2$.}
 \end{center}
\end{figure}

\section{Results}
In FIG. 2 we present a plot of the $\chi^2$ as a function of the best-fit $H_0$ for uncoupled models, where a dashed line denotes the $\chi^2$ for the $\Lambda$CDM model, $\alpha = \beta = \gamma = 0$. From the plot it is clear that deviations from $\Lambda$CDM that lower $H_0$ are favoured by the data. In contrast, $H_0$ may increase, but this worsens the fit to the data. Interestingly, we find no exceptions to this statement in {87,675} random models and  we have checked that adding a $H_0$ prior \cite{Riess:2019cxk} throughout does not change our conclusions.

\begin{figure}[htb]
\begin{center} 
    \includegraphics[scale=0.6]{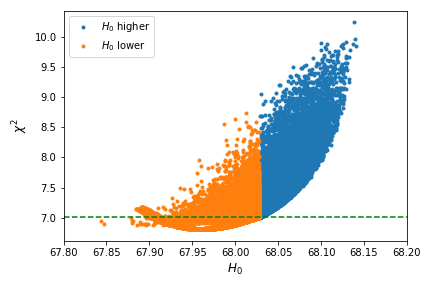}
   \caption{Distribution of {80,732} surviving models as a function of $\chi^2$ versus $H_0$ for the coupling $C=0.1$. Here all models that raise $H_0$ also raise the $\chi^2$, whereas {46,665} models lower $H_0$ and $\chi^2$.}
 \end{center}
\end{figure}

In FIG. 3 we present the same plot but for coupled Quintessence models with $C=0.1$. We are again making a comparison to the $\Lambda$CDM model. Concretely, here we first identified triples of $(\alpha, \beta, \gamma)$, i. e. potentials where $w > -1$, thus ensuring that the uncoupled model is a \textit{bona fide} Quintessence model, before adding the coupling. Once the coupling is added, we can enter the $ w < -1$ region of parameter space and we repeat the fitting procedure. Thus, the configurations presented in FIG. 3 are a subset of the FIG. 2 configurations, but subject to a coupling.  {In contrast to the uncoupled case, the range of $H_0$ is extended}, but since $C$ is small, the effect is not so great. One could choose a larger coupling, but this would bring us into conflict with \cite{Desmond:2018euk, Desmond:2018sdy}. {Finally, we have checked that removing the poorer quality cosmic chronometer data does not change our conclusions.}

Finally, we can make some remarks on the universal constants $c, c'$ from the de Sitter Conjecture \cite{Obied:2018sgi, Garg:2018reu, Ooguri:2018wrx}. We find that models that lower $H_0$ occur at smaller values of $| V'|/V$ relative to models that raise $H_0$. Ostensibly, this means that the de Sitter conjecture may (theoretically) favour models that raise $H_0$. That being said, such models have exclusively $V'' > 0$, whereas models that lower $H_0$ allow $V'' < 0$. For this reason, the Refined de Sitter Conjecture \cite{Garg:2018reu, Ooguri:2018wrx} favours models with lower $H_0$ than $\Lambda$CDM. Note, this is a statement that does not rely on $\chi^2$, but still makes use of fits to data. 

\section{Discussion}
With an eye on the de Sitter Swampland conjecture \cite{Obied:2018sgi}, we have performed a scan over (coupled) Quintessence models at low redshift. Since $\alpha, \beta, \gamma$ are in one-to-one correspondence with $V_0, V'_0, V''_0$, our best-fit values for $H_0$ probe different models (potentials) in the vicinity of $\Lambda$CDM. Within our assumptions and the data employed we arrive at the conclusion that $H_0$ decreases relative to $\Lambda$CDM in any Quintessence model with an exponential coupling. Such models are expected to arise from string theory constructions, e. g.  \cite{Anchordoqui:2019amx, Anchordoqui:2020sqo}, but explicit constructions are especially challenging \cite{Cicoli:2018kdo, Hebecker:2019csg}. 

Our analysis here {employs Taylor expansion around $z=0$, so we are simply assuming analyticity. This allows us to describe a large class of single field Quintessence models with a canonically normalised kinetic term. Nevertheless,} to turn any of these models into viable cosmologies, one needs to ensure that there exists a high redshift completion to CMB. In fact, modulo the fact that truncating the $w_0w_a$CDM or CPL model \cite{Chevallier:2000qy, Linder:2002et} at linear order in $(1-a)$ is guaranteed to constrain the potential (see comments in the appendix) \footnote{Also note that CPL is a linear model and does not allow wiggles in the EoS, whereas coupled Quintessence models do.}, the results of \cite{Vagnozzi:2018jhn} suggest little will change for uncoupled Quintessence models that admit a high redshift completion to CMB. So, uncoupled Quintessence is clearly at odds with high local determinations of $H_0$ \cite{Riess:2019cxk, Wong:2019kwg, Freedman:2019jwv, Pesce:2020xfe}. In short, if Hubble tension in $\Lambda$CDM is now a ``problem" or ``crisis", Quintessence simply instills a sense of panic. 

This leaves a coupling between Quintessence and dark matter as a potential loophole. Here we have focused on the exponential coupling, which is the most natural from string theory, and imposed strict bounds that are derived from the absence of differential fifth forces in galaxy warps \cite{Desmond:2018euk, Desmond:2018sdy}. We find that the coupling is too small to induce any effect within the redshift range. Indeed, it is clear from {equation (\ref{h})} that the $O(z)$ term in the Hubble parameter is unaffected by the coupling, so the coupling is suppressed near $z \approx 0$. With such a small coupling, any interaction between Quintessence and dark matter will have to act over an  extended redshift range, in other words beyond the dark energy-dominated regime, to make progress on Hubble tension. In particular, we may have to embrace a coupling that acts in the dark ages \cite{Agrawal:2019dlm}, which is currently beyond the scope of experiment, but interestingly, this may quickly change \cite{Bowman:2018yin}.

\section*{Acknowledgements}
We thank S. Angus, S. Appleby, H. Desmond, A. Hebecker, C. Krishnan, A. A. Sen, P. Steinhardt, S. Vagnozzi, C. van de Bruck and J. Zhang for correspondence and discussion. This work was supported in part by the Korea Ministry of Science, ICT \& Future Planning, Gyeongsangbuk-do and Pohang City. LH is supported by funding from the European Research Council (ERC) under the European Unions Horizon 2020 research and innovation programme grant agreement No 801781 and by the Swiss National Science Foundation grant 179740. MMShJ would like to thank the hospitality of ICTP HECAP  where this research carried out. MMShJ acknowledges the support by 
INSF grant No 950124 and Saramadan grant No. ISEF/M/98204.

\appendix

\section{Comment on CPL model}
{In this section, we review the analysis of Scherrer \cite{Scherrer:2015tra}, before taking it further using a combination of CMB, BAO and SNe data. Recall that the CPL model \cite{Chevallier:2000qy, Linder:2002et} can be expected to correspond to a specific class of potentials when $w(z) \geq -1$ \cite{Scherrer:2015tra}. More concretely, one can solve the potential in terms of the scale factor \cite{Scherrer:2015tra}, 
\be
V(a) = V_0 \frac{(1- w_0 -w_a + w_a a)}{(1-w_0)} a^{-3 (1 + w_0 +w_a)} e^{3 w_a (a-1)}, 
\ee
where we have normalised the potential so that $V(a=1) = V_0$. Now, to determine $\phi$ as a function of $a$, one just needs to perform the following integral: 
\be
\label{phi_eq}
\phi = \int \frac{\sqrt{3 (1+ w_0 +w_a - w_a a)}}{\sqrt{1 + (\rho_{m0}/\rho_{\phi 0}) a^{3(w_0+w_a)} e^{3 w_a (1-a)}}} \frac{\dd a}{a}, 
\ee
where following \cite{Scherrer:2015tra}, we adopt $\rho_{m0}/\rho_{\phi 0} = 3/7$. Note that $\phi$ is only defined up to a constant, but to fix it we assume $\phi(a=1) = 0$. As can be seen from FIG. 1 or FIG. 2 of \cite{Scherrer:2015tra}, CPL models are not generic and they correspond to a specific class of Quintessence models. }

{Here, we will take the observation a little further in light of the results of Vagnozzi et al. \cite{Vagnozzi:2018jhn}. There, given that they are fitting CMB data at $z \approx 1090$, the authors impose the conditions, 
\be
\label{condition}
w_0 \geq -1, \quad w_0 + w_a \geq -1, 
\ee
which will guarantee that $w(z) \geq -1$ for all $z$. So, here we will take the CMB+BAO+Pantheon MCMC chains from Planck analysis of the CPL model \cite{Aghanim:2018eyx} and impose the constraints (\ref{condition}). This then leaves one with 400 odd pairs of $(w_0, w_a)$, which given the preference of the data for the cosmological constant are all pretty close to $(-1, 0)$. We next integrate (\ref{phi_eq}) from $a =  0.00092$ $(z = 1090)$ to $a=1$ ($z=0$). This provides us a mapping between $\phi$ and $a$ for each CPL model, i. e. each $(w_0, w_a)$ pair. One can then work outwards from $\phi = 0$, corresponding to $a=1$, to determine $V/V_0$ as a function of $\phi$. In practice, we bin $V/V_0$ into bins of $\Delta \phi = 0.01$ and extract out the mean and $1 \, \sigma$, $2 \, \sigma$ intervals in each bin, before applying a spline to these values. The end result is shown in FIG. 4.}

{There are a number of take-homes from the plot. First, for reasons that we have explained above, the Quintessence potential corresponding to the CPL model is pretty constrained. This is even more apparent when one integrates (\ref{phi_eq}) all the way to $a \approx 0$. This is in line with earlier comments in ref. \cite{Scherrer:2015tra}. Secondly, CMB+BAO+Pantheon data has a clear preference for the cosmological constant $\Lambda$, so it is no surprise that the potential is extremely flat as one approaches $\phi = 0$ ($a =1$). Finally, it is worth noting that if one interprets this even in terms of the simplest Quintessence model, one may come to the erroneous conclusion that they are ruled out. This is not the case, it is just simply that the Quintessence potentials corresponding to the CPL model are indeed very special, so special that $V = V_0 e^{- \lambda \phi}$ falls outside the class for certain ranges of $\phi$, or the corresponding ranges in $a$. That being said, it should be easy to show that our results are consistent with \cite{Heisenberg:2018yae} in the small $\phi$ regime, i. e. at late times.}

\begin{figure}[htb]
\begin{center} 
    \includegraphics[scale=0.6]{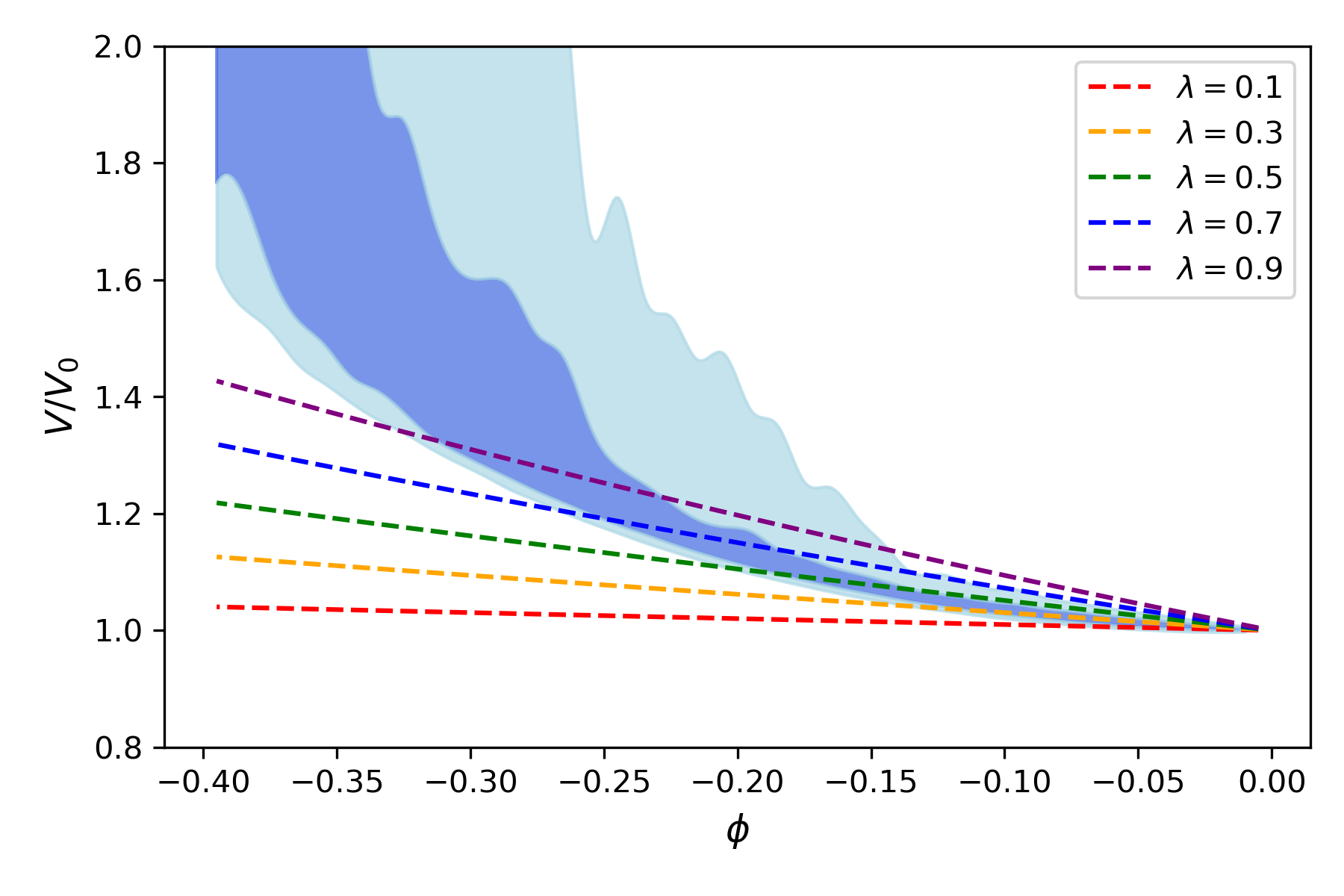}
   \caption{The $1$ and $2 \, \sigma$ intervals for the normalised potential $V/V_0$ corresponding to CPL models that are consistent with CMB, BAO and Pantheon. For illustrative processes we plot the simplest Quintessence model, $V/V_0 = e^{-\lambda \phi}$ for different values of $\lambda$.}
 \end{center}
\end{figure}

\end{document}